\documentclass[12pt]{article}
\usepackage{amssymb}
\usepackage{graphicx}
\begin{document}
\title{Quantum transition  in
bilayer states}
\author {  Vincent Pasquier.}
\date{}
\maketitle
\hskip-6mm
{ Service de Physique Th\'eorique, C.E. Saclay,
91191 Gif-sur-Yvette, France.}

\begin{abstract}
  I study the possible phase transitions when two layers at filling
  factor $\nu_t=1$ are gradually separated. In the bosonic case the
  system should undergo a pairing transition from a Fermi liquid to an
  incompressible state. In the Fermionic case, the state evolves from
  an incompressible $(1,1,1)$ state to a Fermi liquid. I speculate
  that there is an intermediate phase involving charge two
  quasiparticles.

\end{abstract}
\maketitle

\section{Introduction}
The quantum Hall effect \cite{PRA} is both a quantum and a macroscopic
phenomena.  Both aspects manifest themselves through the transport
properties.  The quantum character cannot be understood without
invoking the splitting of levels $ \hbar eB\over m$ induced by the
magnetic field $B$ ($m$ is the mass of the electron).

The relevant parameter which characterizes the system is its filling
factor $\nu$ related to the electron density in units of magnetic flux
(typically $10^{10}$ electrons per square cm for magnetic fields
of a few Tesla).  For a small density $\nu<<1$ the electrons form a
crystal due to the quenching of the kinetic energy. Experiments have
shown that the system is a liquid which conducts the current up to
quite small filling factors ($\nu \sim 1/7$).  Moreover, the
conductivity tensor
\begin{eqnarray}
\pmatrix   {\sigma_{xx}\  \sigma_{xy} \cr
  \sigma_{xy}\  \sigma_{yy}  }
\label {TEN}
\end{eqnarray}
has very peculiar features: $\sigma_{xy}$ is strictly constant and
equal to $\nu e^2/h$ with $\nu$ a fractional filling factor for a
wide variation of the magnetic field called the plateau region. It
increases rapidly to reach a higher fractional value in between two
plateaus. In the plateau regions $\sigma_{xx}$ is strictly equal to
zero and suddenly grows to reach large values in between the plateaus.
Each plateau corresponds to a phase characterized by a specific wave
function for the ground state. The system develops a gap responsible
for the vanishing of the dissipative conductivity $\sigma_{xx}$.
The transition region where the system
switches between two plateaus is the quantum analogous of a continuous
phase transition.

Here we investigate other types of transitions which occur when two
electron (or bosonic) layers are separated from each other.  In this
case, the filling factor is kept fixed and the continuously varying
parameter is the separation $d$ between the two layers. The quantum
transition results from the weakening of the interlayer interactions
as they are separated. Two phases with a definite wave function can be
identified when the layers are either very close or very far from each
other

We consider cases where the total filling factor is less than one and
the dynamics is restricted to the lowest Landau level. The way
particles organize is counter intuitive because their position is not
a good quantum number any more. Instead, we must use the guiding
center momentum $P_x,P_y$ to localize them. In the symmetric gauge,
for example, the expression for $P_x,P_y$ are given by:
\begin{eqnarray}
P_x=p_x-qy/2,\ P_y=p_y+qx/2
\label {MOM}
\end{eqnarray}
These guiding center coordinates do not commute: $[P_x ,P_y]=-iq$ where
$q= eB/{\hbar c} $ is the charge of the particle times the magnetic field.
As a result one cannot localize a particle better that over a cell
of area $2\pi l^2$ with $l^2=q^{-1}.$ We can imagine that the effect
of the magnetic field is to divide the space into cells, each of which
corresponds to a quantum state.  The precise definition of the filling
factor $\nu$ is the number of electrons per cell.  Note that the mass $m$
is an irrelevant parameter which only appears in the
level splitting and disappears from the dynamics. As a result, all the
relevant parameters are solely due to the interactions.  In principle
it is a degenerate perturbation problem where the effective
Hamiltonian is obtained by projecting the interaction potential $V$ in
the lowest level. If we denote by $P$ this projector, the effective
Hamiltonian is given by:
\begin{eqnarray}
H=PVP
\label {HAM}
\end{eqnarray}
Essentially, the effect of the projection is to replace the
coordinates of the potential by the guiding center coordinates.
Therefore $H$ is a true operator ( it has non diagonal matrix
elements) acting in the LLL Hilbert space.

In the fractional Hall effect, the plateaus can be explained through a
careful study of the dynamics induced by (\ref{HAM}). The aim here is
to analyze similar phenomena in the bilayer systems.  The systems are
made of two layers and switch from one phase to the other as the
separation between the layers is increased. A transition is expected
to occur for $d\sim l$.

I consider the case of electron bilayers of current experimental
\cite{SPI} and theoretical \cite{GRE,KIM,YO} interest. The approach I
follow is very closed to the one of Kim et al. \cite{KIM} (especially
their second section) although some of the conclusions are in better
agreement with the recent proposal of Nomura and Yoshioka \cite{YO}.
I also study bosonic bilayers technically easier to understand which
are potentially observable in the context of rotating Bose condensate.

\section{Exciton in electron bilayers}

The system made by two parallel layers of electrons has attracted a
lot of experimental attention. In particular Spielman {\it et al.}
\cite{SPI} have observed a huge enhancement of the tunneling
conductance at small separation.

When the separation $d$ between the two layers is large they behave
independently and are described by a gapless Fermi liquid \cite{HLR}
On the other hand, as $d$ is reduced, the system undergoes a
transition to an incompressible state described by the filling factor
$\nu_T=1/2+1/2=1$.  Kellog et al.  \cite{SPI1} have clearly exhibited
the strong quantum Hall drag resistance which sets up in this regime.
In this section I only discuss the incompressible state obtained when
the layers are very close $d<l$ and live the description of the
compressible state and the transition to a further section.

Consider the case where both layers are on top of each other.
Electrons in one layer are pseudospin up while those in the other
layer are pseudospin down. The system must be in a ferromagnetic state
if we assume that the effect of the interactions can be reduced to a
short range repulsive potential. In the symmetric gauge the spatial
part of the wave function for $N_e$ electrons is then equal to a
Vandermonde determinant, the so-called $(1,1,1)$ state in Halperin's
terminology \cite{HALP}:
\begin{eqnarray}
\Psi_{1,1,1}=\prod_{i<j}
( z_i^{\uparrow}-z_j^{\uparrow})
 (z_i^{\downarrow}-z_j^{\downarrow})
( z_i^{\uparrow}-z_j^{\downarrow})
\label {111}
\end{eqnarray}
The three $1$ in $(1,1,1)$  refer to the exponents of each of the
three facors
in (\ref{111}).
It is the unique wave function at
filling factor one which vanishes when any two electrons are at the same
position.  The Pauli principle then forces the pseudospin part to be
symmetric and therefore the pseudospin takes its maximum value
$N_e/2$. If both layers are exactly half filled, one has
$N_{\uparrow}=N_{\downarrow}=N_e$ so that the $z$ component of the
pseudospin $N_{\uparrow}-N_{\downarrow}$ is equal to zero and the
pseudospin points in the $x-y$ easy-plane. The natural excitations are
spin waves with a quadratic dispersion relation characteristic of a
ferromagnet and not a linearly dispersing Goldstone collective mode.
Said differently, the groundstate is a condensate of excitons obtained
by acting with $(S^{-})^{N_{\downarrow}}$ on the state with all
electrons in the top layer.

This description can be refined using an excitonic picture.  The
excitons can be introduced by starting from a situation where the
pseudospin up layer is filled and the other layer is empty. Suppose
one electron is removed from the top layer to be put in the down
layer. In this simple situation one has a hole in the up-layer
interacting with an electron in the down-layer. The dynamics can be
solved using the model Hamiltonian (\ref{HAM}) and it can be shown
that the electron and the hole form a bound state \cite{KAL}. The main
point of the following discussion is to show that the bound state wave
function is independent of the the interacting potential and has a
group theoretical interpretation \cite{PAS}.  Particles in the lowest
Landau level organize into representations of a deformation of the
displacement group generated by $P_x,\ P_y$ (\ref{MOM}) and the
angular momentum $L=xp_y-yp_x$ obeying the relations:
\begin{eqnarray}
[P_x,P_y]=-iq,\ [L,P_x]=iP_y,\ [L,P_y]=-iP_x
\label {REL}
\end{eqnarray}

They are characterized by the charge $\pm q$ of the particle (in
(\ref{REL}) the charge $q$ must be replaced by $-q$ when we consider a
hole)  which play a similar role as the angular momentum for the
rotations ( the Casimir operator is $P_x^2+P_y^2-2qL$).  The wave
function of the bound state is the analogous of a Clebsh-Gordon
coefficient which couples the two representations of charge $ q$ and
$-q$ into an irreducible representation of charge zero. The exciton
having a zero charge, it does not feel the exterior magnetic field and
its guiding center coordinates $P_x,\ P_y$ can be diagonalized
simultaneously.  The bound state is a dipole oriented perpendicularly
to its momentum ${\bf P}$ of length $Pl^2$.  The dispersion relation
can be computed in terms of the interaction $V$ and is quadratic at
low momentum $\epsilon(p)\sim V p^2l^2/2$.  Excitons behave as
effective bosons interacting with the Hamiltonian \cite{PAHA}:
\begin{eqnarray}
H={1\over 2\Omega}
\sum_{a,b}\int V_{ab}(x-y) \rho_a(x)\rho_b(y) d^2x d^2y
\label {HAMI}
\end{eqnarray}
Here $a,b$ is a layer index and the Hamiltonian takes into account the
fact that the attraction between different layers is weaker than the
repulsion in the same layer:
$V_{\uparrow,\downarrow}<V_{\uparrow,\uparrow}$. Note that
(\ref{HAMI}) is nothing but the second quantized rewriting of
(\ref{HAM}).  When the two layers are on top of each other ($d=0$),
the $SU(2)$ symmetry is recovered and the excitons interact weakly, which
explains why the dispersion relation is quadratic in the momentum. The 
first effect of the separation is to introduce a repulsion
between the excitons. If we model them by a slightly non-ideal
Bose gas, the dispersion relation is parameterized by the repulsion
pseudopotential $U_0$ equal to zero for $d=0$ and
increasing with $d$.  As a result the exciton behaves like a Goldstone
boson with a sound velocity $u\sim \sqrt{U_0/ml^2}$ increasing
with the separation. This goes with a smooth decreasing of the total
spin as seen in \cite{YO}. The apparent contradiction between the absence
of a gap and the observed incompressibility is due to the fact that
the exciton is neutral and does not interfere with the charge gap
responsible for the Hall effect.

When the separation $d$ is so large that $V_{\uparrow,\downarrow}=0$
the fluid is no more incompressible and the Bose-gas picture is not
correct any more. Instead, each layer can be modeled by a gas of
neutral fermionic dipoles \cite{READ,PAHA,DIP}. The problem is then to 
understand how the transition between the gas of bosonic excitons
and the two uncorrelated Fermi liquids occurs.

\section{boson bilayers}

Before discussing this quantum transition, I wish to draw an analogy
with the reversed phenomena that occurs when two bosonic layers are
moved away from each other.  A physical context could be two
Bose-condensate in a rotating trap gradually separated from each
other. To use a language adapted to the Hall effect, I treat the
rotation as if it were a magnetic field which means that the rotation
frequency is equal to the harmonic trap frequency \cite{WIL}. The
magnetic length is then defined in terms of this critical frequency.

The physical problem consists of two kinds of bosons in a magnetic
field at filling factor $\nu=1$. We imagine that the particle index is
a layer index.  At zero separation the interaction between particles
in different layers is the same as the interaction between particles
in the same layer.  By analogy with the quantum Hall state at
$\nu=1/2$, we expect the system to be described by a Fermi liquid
state. We then separate the two layers which are exactly at half
filling.  When they are sufficiently far away that particles between
different layers do not interact any more, one is left with a two
copies of a $\nu=1/2$ bosonic system which are incompressible states.
Therefore, we expect that a transition will occur at some separation
where the dissipative conductivity suddenly vanishes as the
incompressible state builds up.  This is exactly the reverse situation
as with electrons, one goes from a compressible Fermi liquid state to
an incompressible boson condensate as the two layers are separated
from each other.

Let us first consider the zero separation state which should
correspond to a Fermi liquid state.  The picture developed for the
Fermi liquid state at $\nu=1$ \cite{PAHA} is in terms of neutral
fermionic dipoles consisting of the charge $e$ boson and a fermionic
hole with a charge $-e$.  Energetically, the system would like to have
a slater determinant wave function (\ref{111}). This wave function is
however in conflict with the bosonic statistics and the neutral
dipole Fermi liquid is the less costly manner in which it adjusts
itself to satisfy the correct statistics. It can be written in a
product form \cite{REZ}:
\begin{eqnarray}
\Psi(z_i)=P\{ {\rm Fermi\ Liquid}\ \prod_{i<j}z_i-z_j \}
\label {FLW}
\end{eqnarray}
As explained earlier, the projection $P$ is the origin of the dipole
interpretation of the Fermionic quasiparticles. Each Fermi liquid
quasiparticle is a dipole made of a boson correlated to a hole in the
Slater determinant factor. The difference between
the dipoles and the bilayer excitons is their fermionic statistics.

Another possibility to satisfy the Bose statistics is the Pfaffian
state \cite{MOO}:
\begin{eqnarray}
\Psi(z_i)={\rm Pf}\{{1 \over z_i-z_j}\}\prod_{i<j}z_i-z_j 
\label {PFA}
\end{eqnarray}
The Pfaffian factor being antisymmetric guarantees that the total wave
function is symmetric. Each denominator $1/ z_i-z_j$ in the Pfaffian
removes the correlation hole between particles $i$ and $j$. Thus, the
Pfaffian induces a pairing between the particle and can be thought as
a kind of BCS wave function where the composite Fermions are in a 
p-paired $L=-1$ state. Unlike the Fermi liquid, this state
is incompressible.

In the bilayer case the boson carry a spin index which specifies in
which layer they lie. The Fermi liquid takes advantage of this to
reduce its energy by putting two dipoles with up and down spin in the
same momentum state thus reducing the Fermi momentum by a factor
$\sqrt{2}$ with respect to the spinless case. The system is in the
paramagnetic state. The Pfaffian state on the other hand is
ferromagnetic and can be obtained by acting with
$(S^-)^{N_{\downarrow}}$ on the state with all bosons in the top layer
without energy gain. As a result the Fermi liquid is probably
energetically favored with respect to the Pfaffian in this bilayer
situation.

In the large separation limit limit the $\nu=1/2$ bosonic state of one
layers is a $\nu=1/2$ Laughlin type wave
function
\begin{eqnarray}
\Psi(z_i)=\prod_{i<j}^{N/2} (z_i-z_j)^2
\label {PAIRI}
\end{eqnarray}
which is legitimate for bosons. This state is incompressible and
minimizes the energy of a single layer. To understand how the
transition from a Fermi liquid to this kind of state occurs it is
useful to rewrite the product $(2,2,0)$ of the two $\nu=1/2$ factors
(\ref{PAIRI}) as a paired state:
\begin{eqnarray}
\Psi(z_i)={\rm Det}\{{1 \over z_i^{\uparrow}-z_j^{\downarrow}}\}
\prod_{i<j}^{N}z_i-z_j 
\label {PAIR}
\end{eqnarray}
which can be shown using the Cauchy identity. This rewriting clearly
shows that the large separation limit can be understood as a pairing
between the bosons of the top layer with those of the bottom layer.
The pairing factor ${1 / z_i^{\uparrow}-z_j^{\downarrow}}$ annihilates a
correlation hole between these two bosons and carries an angular
momentum $L=-1$. 

If the state at zero separation is the Pfaffian state (\ref{PFA}), it
can be continuously deformed \cite{HO} into the state (\ref{PAIRI})
without undergoing a phase transition. For this one needs to multiply
the matrix element in (\ref{PFA}) by a factor $1+\mu \sigma_i\sigma_j
$ and let $\mu$ vary between $0$ and $-1$.

The pairing instability also follows in the dipole approach \cite{KIM}.
The dipoles at the Fermi surface have a length $k_fl^2$ and an
orientation perpendicular to their momentum. For obvious geometrical
reasons a dipole with momentum ${\bf k}$ tends to bind with a dipole
${\bf -k}$. When the repulsion between bosons of the two layers
decreases, this strengthens the binding between dipoles with opposite
spins and very plausibly induces the pairing instability in the
p-channel.

To conclude this section, two scenarios are possible in the case of
bosonic bilayer systems. In the first one the system is
incompressible at all separations and it is in the Pfaffian state at
zero separation. In the second more probable one the state is a
Fermi liquid at zero separation and undergoes a pairing transition to
an incompressible state as the separation is increased.

\section{Transition in electron bilayers}
I now return to the transition in the $\nu_t=1$ Fermionic layers. The
problem is more difficult and this section is speculative.

The intuition gained in the bosonic bilayer case was that layer
separation induced attraction between the bosons in different layers
at $d\ne 0$ which resulted in the disappearance of the correlation
hole between them. What made life easy was that the quasiparticles
relevant at $d=0$ were Fermions and the pairing mechanism was
reminiscent of a BCS transition. In the present case, excitons are
bosons and we have seen that the effect of the separation is to repel
them.  Therefore we abandon the exciton picture and try to model the
transition as a pairing mechanism between the electrons directly.
This is possible if we multiply the wave function (\ref{111}) by a
symmetric factor which does not spoil its polynomial character nor
modifies the filling factor. It suggests to multiply the 
wave function by a Permanent Factor \cite{HR}:
\begin{eqnarray}
\Psi(z_i)={\rm Per}\{{1 \over z_i^{\uparrow}-z_j^{\downarrow}}\}
\Psi_{1,1,1}
%=|{1 \over (z_i^{\uparrow}-z_j^{\downarrow})^2}|
={\rm Det}\{{1 \over (z_i^{\uparrow}-z_j^{\downarrow})^2}\}
\prod_{i,j} (z_i^{\uparrow}-z_j^{\downarrow})^2 
%\nonumber
\label {HR}
\end{eqnarray}
The second equality results from Borchart identity \cite{GAUD}.  The
first writing exhibits it can be obtained as a paired wave function
from the state $(1,1,1)$ (\ref{111}). The second writing represents
the state as a paired state built on the $(0,0,2)$ bosonic Laughlin
state.  Note that the weak pairing state with the square factor
removed in the determinant yields back the $(1,1,1)$ state \cite{KIM}.
Although the two states (\ref{111}) and (\ref{HR}) are both
incompressible, the transition should have consequences in drag
experiments \cite{SPI1}.  In (\ref{111}) electrons of the first layer
are bound to holes in the second layer whereas in (\ref{HR}) they form
pairs with the electrons of the second layer in agreement with the
conclusions of \cite{YO}. 

This cannot be the complete story however since at large separation
the Fermi liquid states are built as in (\ref{FLW}) on a $(2,2,0)$
incompressible state, not a $(0,0,2)$ one as in (\ref{HR}).  A
possible precursor to the Fermi liquid state is a product of two Pfaffian
states. The main difference between the Pfaffian states and our trial
state (\ref{HR}) is that in the Pfaffian the electrons are paired
inside one single layer whereas in (\ref{HR}) the pairs involve two
electrons in different layers. It is possible that this repairing
occurs in a continuous way. The Pfaffian incompressible state then
undergoes a second phase transition towards a Fermi liquid state.
Although this scenario with two phase transitions is neither
economical nor easy to formalize precisely, it is difficult to rule
out an intermediate phase involving paired quasiparticles. An
experimental compelling evidence of this possibility would be to
observe charge two carriers in this intermediate phase.

\section{Concluding remarks}
The two layer systems clearly exhibit quantum phase transitions
mediated by interactions. Such transitions are now well studied in the
electron context and it would be very interesting to see them in
bosonic systems. A first step would however be to clearly identify a
fractional Hall regime in rotating Bose condensates and the most
promising direction seems to me the analogous of the Jain series at
filling factors $\nu=p/p+1$ terminating in a Fermi liquid state at
$\nu=1$.  The transition discussed here would be a second step.  In
the fractional Hall regime, the message of this essay is to stress
that the most likely transition between a Fermi liquid state and an
incompressible state is through a pairing mechanism which may
hopefully be seen experimentally.

\section*{Acknowledgements}
I thank Christian Glattli for persuading me to write this essay
and Jean Dalibard for drawing my attention to the
problem of phase transitions in Bose condensates.

\end{document}